\documentclass[10pt, twocolumn]{article}

\usepackage[utf8]{inputenc}
\usepackage{microtype} % Dramatically improves text justification and hyphenation
\usepackage{mathpazo} % Uses the elegant Palatino font for a classic whitepaper look
\usepackage{titlesec} % For custom section headers
\usepackage{xcolor} % For subtle color accents
\usepackage{geometry}
\usepackage{fancyhdr}
\usepackage{lettrine} % For the sleek drop-cap at the start
\usepackage{amsmath}
\usepackage{amssymb}
\usepackage{graphicx}
\usepackage[colorlinks=true, linkcolor=cyan!80!black, urlcolor=cyan!80!black]{hyperref}

\renewcommand{\headrulewidth}{0.5pt}
\renewcommand{\headrule}{\hbox to\headwidth{\color{black!30}\leaders\hrule height \headrulewidth\hfill}}
\titleformat{\section}{\large\bfseries\scshape\color{black!90}}{\thesection.}{0.5em}{}
\titleformat{\subsection}{\normalsize\bfseries\color{black!80}}{\thesubsection.}{0.5em}{}

\title{\vspace{-1.5cm}\huge\bfseries The Structural Case for the Eco-Civilization Paradigm \\ \vspace{0.2cm} \Large Thermodynamic Limits, Asset Valuation, and Decentralized Infrastructure}
\author{\scshape Lei Zhu \, \, \, \scshape William Zhu}
\date{\vspace{-0.5cm}} 

\begin{document}

\maketitle
\begin{abstract}
\noindent This study establishes a quantitative and structural framework for civilizational continuity under rapid, non-linear ecological transitions. Utilizing empirical data on Earth’s Energy Imbalance (EEI), currently measured at $1.36\text{ W/m}^2$, we demonstrate that the planetary system is on a deterministic trajectory toward a new ecological equilibrium, breaching the $1.5^\circ\text{C}$ thermodynamic threshold in less than 6.5 years. We show that mitigating direct anthropogenic waste heat emissions is statistically insufficient, as $35/36\text{-ths}$ of the annual energy imbalance stems from atmospheric radiative forcing rather than localized thermal output. 

Consequently, preserving legacy centralized socioeconomic frameworks is unfeasible. We present the structural case for the ``Eco-Civilization Paradigm''---a decentralized network of autonomous, modular eco-communities engineered to insulate core life-support functions, localized energy networks, and edge computation from cascading macro-system failures. Finally, we outline the systemic requirements for this transition, including the realignment of asset valuation frameworks from geography-dependent to engineering-dependent metrics, the mobilization of youth-driven innovation capital, and the strategic reallocation of global industrial manufacturing capacity to safeguard civilizational survival.
\end{abstract}
\vspace{0.5cm}

\section{The Timeline to a New Ecological State}

\lettrine[lines=2, lhang=0.33, nindent=0em]{E}{arth} is on a measurable trajectory toward a new ecological equilibrium \cite{zhu2026planetary1, zhu2026planetary2}. The Earth's Energy Imbalance (EEI) is currently $1.36\text{ W/m}^2$, representing the net operational difference between incoming solar radiation and outgoing longwave radiation at the top of the atmosphere \cite{hansen2023global}. This imbalance results in an annual accumulation of approximately $2.19 \times 10^{22}\text{ joules}$ of waste heat.

Empirical measurement systems indicate that approximately 90\% of this excess energy is sequestered within the global oceans, driving a steady increase in ocean heat content, accelerating marine thermal expansion, and altering deep-sea currents \cite{vonschuckmann2020heat}. The remaining 10\% is distributed across the atmosphere, continental landmasses, and the cryosphere, where it accelerates the ablation of ice sheets and glaciers.

The remaining margin before reaching an irreversible $1.5^\circ\text{C}$ global temperature increase is $1.42 \times 10^{23}\text{ joules}$. This specific capital budget is determined by the calculated heat capacity of the troposphere and upper ocean layers relative to the baseline of pre-industrial temperatures \cite{vonschuckmann2020heat}. Based on current data and projections, this critical thermodynamic threshold threats to trigger an irreversible cascade of planetary system changes in less than 6.5 years \cite{hansen2023global, meng2023domino}.

\section{Quantitative Limits of Anthropogenic Heat Mitigation}

Altering this trajectory through the reduction of direct human heat emissions is unfeasible. Direct physical waste heat generated by human society---defined as the immediate thermal dissipation from industrial manufacturing, power generation facilities, transportation networks, and residential climate control---is approximately $6.0 \times 10^{20}\text{ joules}$ annually, which represents $1/36\text{-th}$ of the total EEI.

This data reveals a critical structural distinction between direct thermodynamic heat injection and indirect radiative forcing \cite{hansen2023global}. The primary cause of the planetary energy accumulation is not the physical heat released by human machinery, but rather the enhanced insulation of the atmosphere. This insulation traps incoming solar energy that would otherwise radiate back into space. Because $35/36\text{-ths}$ of the annual energy imbalance stems from this solar retention mechanism rather than localized thermal output, current initiatives focused strictly on mitigating urban heat islands or lowering the thermal footprint of mechanical processes are statistically insufficient to alter the macro-trend. Even a hypothetical complete cessation of all direct human thermal emissions would leave the overarching planetary accumulation baseline virtually unchanged, failing to extend the transition timeline.

\section{The Structural Case for the Eco-Civilization Paradigm}

The transition of the biosphere into a new ecological state poses severe structural challenges to centralized models of human organization. Transitioning toward decentralized, autonomous ``eco-civilization communities'' serves as a primary framework for maintaining systemic continuity. When a highly integrated network undergoes a significant phase shift, maintaining rigid, centralized systems increases the probability of widespread operational disruptions. Under these conditions, modularity becomes a practical requirement for stability.

\subsection{Structural Limitations of Centralized Adaptation}
Modern civilization relies on a tightly coupled, integrated global network. This specific topology presents distinct vulnerabilities when facing sustained ecological volatility:
\begin{itemize}
    \item \textbf{Risk of Cascading Failures:} Current global infrastructure depends on centralized hubs for food production (agricultural breadbaskets), energy distribution (interconnected grids), and data processing (remote data centers). In a tightly coupled network, a severe disruption at a critical node can propagate across the entire system, leading to concurrent disruptions in logistics, financial markets, and resource distribution.
\end{itemize}

\subsection{The Geopolitical Unfeasibility of Mass Relocation}
Historically, human societies mitigated regional environmental stress through nomadic migration to more favorable geographic regions. However, simply moving displaced populations to newly suitable or stabilizing environments is no longer a viable option. Modern geopolitical realities---defined by rigid national borders, sovereign immigration controls, and international legal constraints---prevent large-scale population reallocation across territories. Because cross-border mass migration is geopolitically unfeasible, civilizational stability cannot be achieved by attempting to move populations to a better environment; instead, a sustainable environment must be engineered and maintained locally where people already reside.

\subsection{Historical Correlations Between Ecological Stress and Social Disruption}
Empirical historical data indicates that relying strictly on volatile open-air systems during periods of rapid ecological change has consistently led to structural decline:
\begin{itemize}
    \item \textbf{The Resource-Conflict Sequence:} Paleoclimatic records demonstrate that when complex, centralized societies---such as the Akkadian Empire, the Late Bronze Age states, or the Classic Maya---experienced prolonged environmental shifts, they frequently faced a predictable sequence: agricultural decline, institutional strain, localized scarcity, and an increased frequency of internal or external conflict.
    \item \textbf{Modern Systemic Interdependence:} Because modern societies are highly resource-dependent and interconnected, attempting to sustain existing open-air economic structures without adaptive insulation increases the probability of similar disruptive outcomes on a broader scale.
\end{itemize}

\subsection{Modularity as a Resilience Mechanism}
Given that regional environments are projected to exhibit non-linear volatility during the transition phase, maintaining systemic stability supports the adoption of the eco-civilization paradigm.

Network analysis demonstrates that modular systems manage shocks more effectively by isolating localized failures rather than allowing them to propagate. By decoupling core life-support functions---such as food production, energy generation, and computational resources---from global distribution networks, eco-civilization communities function as resilient, autonomous units. This approach shifts the baseline of human organization from a vulnerable dependence on open-air stability to a distributed architecture capable of maintaining internal equilibrium.

\subsection{Possible Path of Emergence}
The deployment of distributed eco-communities will follow a distinct geographic and chronological trajectory based on regional environmental pressures. The initial establishment of these networks will occur within territories experiencing the earliest and most severe ecological disruptions, where traditional open-air infrastructure fails first.

As environmental instability expands into additional territories, the proliferation of these communities will scale alongside the continuous technical iteration of their core survival systems. The growth and construction rate of this decentralized infrastructure will accelerate in direct proportion to the severity of the transitional chaos, reaching its peak when environmental volatility is at its highest.

Once this transitional phase concludes, the biosphere will settle into a new ecological equilibrium. Because this new baseline state is projected to be incompatible with traditional open-air lifestyles and conventional modes of agricultural production, legacy cities and conventional settlements will persist only as isolated exceptions in low-impact zones. Consequently, the long-term continuity of human civilization will be fundamentally organized around and defined by the spatial distribution, total quantity, and operational quality of these decentralized eco-communities.

\section{Re-Anchoring the Economic Valuation System}

Allocating capital to these decentralized communities represents a fundamental realignment of asset valuation frameworks rather than an unrecoverable capital expenditure:

\begin{itemize}
    \item \textbf{Current Economic Baselines:} Modern financial systems assume a stable open-air environment. Real estate values depend on geographic and coastal stability; logistics networks rely on predictable weather patterns; and agricultural commodities are anchored to open-air cultivation. Existing financial pricing models, such as Discounted Cash Flow (DCF) analyses and sovereign credit risk frameworks, are structurally predicated on a baseline of climate stationarity. Under this current paradigm, environmental variables are treated as external, episodic shocks rather than systemic determinants of capital asset values.
    \item \textbf{Asset Devaluation Risk:} If environmental stability degrades, the foundational assumptions of these asset classes become invalid, leading to significant structural devaluation. This process is already empirically evident in the contraction of global property insurance and reinsurance markets, where institutional underwriters are actively withdrawing coverage from regions exposed to escalating sea-level rise and wildfire anomalies. When insurance markets contract, the underlying real estate loses its collateral viability, triggering a downward valuation cascade across debt and equity instruments. Similarly, because open-air agriculture faces a non-linear reduction in yield predictability, assets dependent on these systems risk becoming ``stranded assets'' characterized by permanent capital impairment.
    \item \textbf{Future Value Capture:} Developing eco-civilization infrastructure today establishes the physical assets for a future economic valuation system dictated by a fundamentally different set of ecological parameters, integrating a broader and richer array of ecological resource values. Consequently, market valuation will transition from geography-dependent metrics to engineering-dependent metrics. Instead of pricing assets based on spatial location or proximity to traditional open-air resources, financial markets will value infrastructure based on its capacity for autonomous power generation, closed-loop water and waste processing, and structural environmental containment. Capital deployed into distributed microgrids, closed-loop vertical food production, and edge-computing infrastructure functions as a strategic hedge, converting exposed, open-air liabilities into resilient capital structures calibrated to this future framework while opening vast development space for emerging industries.
\end{itemize}

\section{Technological Infrastructure: Decentralized Energy and Computing}

The operational viability of these communities depends on independent, distributed energy and computation nodes. Because regional environmental volatility is projected to disrupt centralized power grids and remote data networks, centralized utilities cannot be relied upon as a baseline.

Centralized infrastructure networks are highly vulnerable to environmental disruptions because their interconnected nature propagates localized failures across vast areas. For example, severe heatwaves physically reduce the electricity-carrying capacity of long-distance transmission lines at the exact moment air conditioning demand peaks, creating systemic overloads. When combined with extreme weather events that physically destroy infrastructure---such as high-velocity winds or wildfires tearing through a single utility corridor---the entire centralized grid can experience sudden, widespread blackouts. 

Centralized data centers face similar vulnerabilities; they require massive, continuous supplies of water for cooling and uninterrupted grid connections to operate. In an environment with frequent water shortages and regional grid failures, relying on centralized cloud computing creates a single point of failure. This risk requires a structural shift toward localized computing systems that run on their own independent power grids.

\subsection{Operational Functions of Localized Artificial Intelligence}
Localized communities require automated systems to manage internal survival operations and adapt to external environmental pressures. Internally, AI acts as the operational core of the community, constantly balancing finite resources like power, water, and automated food production. Because environmental volatility causes sudden spikes in energy demand or abrupt drops in resource availability, human reaction times are too slow to prevent system failures. Localized AI matches resource supply with real-time community needs, eliminating waste and automatically configuring battery storage. 

Beyond real-time matching, AI optimizes resource allocation by forecasting localized supply deficits weeks in advance based on upcoming weather trends, allowing the community to proactively ration supplies and adjust daily usage before an actual shortage occurs. Additionally, the system manages closed-loop recycling networks, tracking and automatically routing waste byproducts across different sectors---such as redirecting wastewater from households or excess heat generated by computing machinery directly into indoor farming systems---to extract maximum utility from every available resource. 

Externally, AI is critical for environmental defense and planning. On-site AI models analyze local sensor data to provide early warnings for severe weather threats, allowing the community to secure its physical infrastructure before an event occurs. Furthermore, these systems track long-term environmental shifts in the surrounding region, predicting how local habitats are changing so the community can proactively modify its defenses and resource strategies. Without this continuous, high-speed automated management, independent communities cannot maintain internal stability against unpredictable external conditions.

Consequently, the production of decentralized power systems and localized computing infrastructure will become a primary sector of industrial and economic expansion during this transition period. Market demand will shift from standard hardware to integrated power and computing packages, making independent local nodes a primary asset class in the industrial sector.

\section{Human Capital and Innovation Iteration}

The primary driver of technological adaptation within these communities is the youth demographic. This demographic cohort represents the human capital capable of rapidly developing, testing, and iterating localized technologies to meet the evolving operational demands of the new ecosystem. From the lens of systems evolutionary dynamics, youth innovation should serve as a primary engine for the technological adaptation and functional iteration of these communities. Thomas Kuhn observed in \textit{The Structure of Scientific Revolutions} that paradigm shifts and radical innovations are typically driven by the younger generation, who remain unentrenched in legacy systems. This capacity allows the youth demographic to rapidly translate their understanding of the future ecological environment into engineering momentum for the continuous improvement of distributed survival systems.

Younger generations possess distinct structural advantages in driving this transition, specifically regarding technical literacy and economic adaptability. Younger demographics show the highest rates of fluency in software development, automated systems, and decentralized engineering protocols, which constitute the foundational systems of autonomous communities. Furthermore, this population operates with minimal institutional inertia. Unlike older demographics, younger individuals generally lack deeply entrenched financial or career investments in legacy open-air industries and real estate. They are structurally incentivized to build a new paradigm because their long-term stability depends on its success, unburdened by legacy frameworks.

Historical data from major industrial and technological shifts confirms that rapid operational transitions are consistently led by younger demographics. For example, during the expansion of the digital and internet revolutions, the workforce that designed, tested, and deployed decentralized network protocols was overwhelmingly composed of young engineers who could adapt without needing to unlearn outdated mechanical systems. Today, this pattern is mirrored in open-source hardware and green-energy innovation networks, where youth participation drives accelerated cycles of technological iteration. In an environment that demands constant, rapid experimentation to survive localized environmental changes, the youth demographic provides the practical agility and technical focus necessary to maintain community resilience.

\section{Transfer of Manufacturing and Construction Capabilities}

The efficient industrial manufacturing systems and large-scale construction capabilities currently established by human society constitute the material foundation for building distributed eco-communities. As the biosphere transitions through the chaos phase toward a new equilibrium, high-capacity nations and enterprises must reallocate these manufacturing resources to support the construction of decentralized infrastructure.

\begin{itemize}
    \item \textbf{Optimizing Cost-Effectiveness:} Leveraging existing, highly optimized supply chains and industrial assembly lines allows humanity to maximize its manufacturing advantages. This efficiency is a critical factor in compressing engineering timelines and accelerating the scaling of eco-communities within a limited time window.
    \item \textbf{Utilizing Existing Organizational Expertise:} Directing existing large-scale engineering organization and modular manufacturing capacity into the deployment of eco-civilization communities provides the immediate material input required for these new survival nodes. This approach prevents the systemic efficiency losses associated with rebuilding foundational industries during a period of ecological instability.
    \item \textbf{Requirements for Civilizational Continuity:} Against the macro-backdrop of declining open-air ecosystems, mobilizing the manufacturing capital of high-capacity economies for this transition is not a traditional aid framework; rather, it is a structural requirement to preserve global human civilization.
\end{itemize}

\section{Theoretical Progression and Consensus Building}

Through a progressive deduction across a three-article series, this study utilizes thermodynamic data and systems evolutionary logic to demonstrate that Earth's ecosystem and current socioeconomic structures cannot maintain the legacy open-air equilibrium. Objective data on Earth’s Energy Imbalance (EEI) demonstrates that the transition to a new ecological state is a measurable certainty on a defined timeline. Because direct anthropogenic thermal emissions represent only a small fraction of this imbalance, initiatives focused strictly on preserving the existing open-air status quo are statistically insufficient to alter the macro-trend. 

Consequently, the continuity of civilization demands that global leadership and social groups abandon their fixation on the legacy baseline. Instead, they must develop a global consensus, mobilize material and technological resources across sectors, and prepare for concrete construction actions toward the distributed eco-civilization paradigm.

This structural reorganization is supported by a clear sequence of systemic requirements:
\begin{enumerate}
    \item \textbf{Spatial and Geopolitical Realities:} Because geopolitical constraints render mass cross-border relocation unfeasible, sustainable environments must be engineered and maintained locally where populations reside.
    \item \textbf{Valuation System Reorganization:} Executing structural adjustments to shift asset valuation models---which are implicitly predicated on Earth's current equilibrium---toward a resilient evaluation framework calibrated to the future chaos phase and new equilibrium state.
    \item \textbf{Technological Infrastructure:} Facing the vulnerabilities of centralized utilities caused by planetary ecological collapse, the integration of independent local energy networks with localized computing nodes is a structural requirement. This requires automated artificial intelligence to manage resources and predict environmental disruptions in real time.
    \item \textbf{Human Capital and Innovation Iteration:} The development and evolution of localized survival systems will be heavily driven by the youth demographic. Their technical literacy, systemic innovation capacity, and low institutional inertia serve as the primary engine driving continuous technological iteration within communities.
    \item \textbf{Empowerment of Manufacturing Capabilities:} The highly efficient industrial manufacturing and large-scale construction capabilities currently held by high-capacity nations and enterprises must be directed toward vulnerable global nodes. Leveraging humanity's manufacturing cost-effectiveness to enhance the scalability of the eco-community model is a critical requirement to safeguard civilizational survival.
\end{enumerate}

Ultimately, the progressive deduction across the three consecutive articles of this study provides a conceptual foundation for a new-phase model of risk management and capital allocation. By establishing a consensus on the construction of eco-communities today and starting the preparation and mobilization, human society can sustain its development within Earth's future ecological equilibrium.

\bibliographystyle{unsrt}
\bibliography{main}

\end{document}